\documentclass{aa}
\usepackage{graphics}
\def\a{{$\alpha$}}

\newcommand{\h}{$^{\rm h}$}
\newcommand{\m}{$^{\rm m}$}
\newcommand{\s}{$^{\rm s}$}
\newcommand{\dd}{$\delta$}
\newcommand{\ha}{\rm H$\alpha$}
\newcommand{\hbeta}{\rm H$\beta$}

\newcommand{\hnii}{{\rm H}$\alpha+[$\ion{N}{ii}$]$}
\newcommand{\nii}{$[$\ion{N}{ii}$]$}
\newcommand{\sii}{$[$\ion{S}{ii}$]$}
\newcommand{\OII}{$[$\ion{O}{ii}$]$}
\newcommand{\oii}{$[$\ion{O}{ii}$]$}
\newcommand{\oi}{$[$\ion{O}{i}$]$}
\newcommand{\oiii}{$[$\ion{O}{iii}$]$}

\newcommand{\et}{et al.}

\newcommand{\fluxa}{$10^{-16}$ erg s$^{-1}$ cm$^{-2}$ arcsec$^{-2}$}

\newcommand{\dens}{\rm cm$^{-3}$}

\newcommand{\sulfur}{[S~{\sc ii}]}
\newcommand{\nitrogen}{[N~{\sc ii}]}
\newcommand{\oxygen}{[O~{\sc iii}]}
 
\newcommand{\HNII}{{\rm H}$\alpha+$[N {\sc ii}]}  
\newcommand{\OIII}{$[$\ion{O}{iii}$]$}
\newcommand{\SII}{$[$\ion{S}{ii}$]$}
\begin{document}

%
\title{A new candidate supernova remnant G 70.5$+$1.9}

\author{F. Mavromatakis\inst{1}
\and P. Boumis\inst{2}
\and J. Meaburn\inst{3}
\and A. Caulet\inst{4}
}
\offprints{F. Mavromatakis,\email{fotis@stef.teicrete.gr}}
\authorrunning{F. Mavromatakis et al.}
\titlerunning{A new candidate supernova remnant G 70.5$+$1.9}
\institute{
Technological Education Institute of Crete, Department of
Sciences, P.O. Box 1939, GR-710 04 Heraklion, Crete, Greece.\\
\email{fotis@stef.teicrete.gr} 
\and Institute of Astronomy \& Astrophysics, National
Observatory of Athens, I. Metaxa \& V. Pavlou, P. Penteli, GR-15236
Athens, Greece.\\
\email{ptb@astro.noa.gr}
\and Jodrell Bank Centre for Astrophysics, University of Manchester,  
Manchester, UK., M13 9PL.\\ 
\email{jmeaburn@jb.man.ac.uk}
\and Department of Astronomy, University of Illinois at
Urbana-Champaign, 1002 West Green Street, Urbana, IL 61801-3080, USA. \\
 \email{caulet@astro.uiuc.edu}}
\date{Received ...... / Accepted .......}

\abstract{A compact complex of line emission filaments in the galactic
plane has the appearance of those expected of an evolved supernova
remnant though non--thermal radio and X--ray emission have not yet
been detected.  This optical emission line region has now been
observed with deep imagery and both low and high--dispersion
spectroscopy. Diagnostic diagrams of the line intensities from the
present spectra and the new kinematical observations both point to a
supernova origin. However, several features of the nebular complex
still require an explanation within this interpretation.
}

\maketitle
\section{Introduction}
During observations of a new candidate supernova remnant (SNR) in the
Cygnus region, close to the known remnant CTB~80, a bright
emission line source was detected (Mavromatakis \& Strom
\cite{mav02}). This source, henceforth to be called G 70.5$+$1.9, is
relatively compact (a few arcminutes wide), its boundaries are rather
sharp and it exhibits a tail of diffuse emission to the east, only seen
in the \oiii\ image. We note here that the southern part of this
structure is visible on the POSS plates. However, no other information
is available on this peculiar, optical line source. Published radio
surveys do not show any strong evidence for non--thermal emission at
the position of the optical source, while the ROSAT All--Sky survey
data do not reveal any excess X--ray emission at this position.  Both
could be expected if the filamentary nebula is an evolved SNR.  In
order to study in more detail the detected structures and try to
understand their origin we performed deep optical, flux calibrated,
imaging observations in the low and medium ionization lines of \hnii,
\sii, \oii\ and \oiii. We have also obtained deep spatially--resolved
low--dispersion long--slit spectra at four different slit positions of
this filamentary structure in an effort to get more information about
the actual physical conditions. Spatially resolved long-slit spectra
along one slit position have also been obtained to reveal any motions
typical of an evolved SNR. Information about the observations and
results are given in Sect. 2, while in Sect. 3 we discuss the nature of
these new filamentary structure.
\section{Observations \& Results}
\subsection{Imaging}
The observations presented in this paper were performed with the 1.3 m
Ritchey--Cretien telescope at Skinakas Observatory, Crete, Greece.
The filamentary nebula, G70.5$+$1.9 was observed on July 09 and 10,
2001 with interference filters isolating the optical emission lines of
\hnii, \sii, \oii, and \oiii.  The 1024 $\times$ 1024 SITe CCD used
during the observations had a pixel size of 24 $\mu$m resulting in a
8\arcmin.5\ $\times$ 8\arcmin.5\ field of view.  Two different
pointing directions were performed in order to cover the area of
emission seen in the wide field images of Mavromatakis \& Strom
(\cite{mav02}).  The first concentrated on the bright filaments in the
west, while the second focuses on the diffuse \oiii\ emission in the
east. All fields were projected to a common origin on the sky and were
subsequently combined to create the final mosaics.  The astrometric
solutions were calculated with IRAF routines and utilized the HST
Guide star catalogue (Lasker \et\ \cite{las99}). The log of the
observations together with the filter characteristics and the exposure
times are given in Table \ref{table1}. All coordinates given in this
work refer to epoch J2000.
\par
We employed standard IRAF and MIDAS routines for the reduction of the
data.  Individual frames were bias subtracted and flat--field
corrected using well exposed twilight flat--fields. The
spectrophotometric standard stars HR5501, HR7950, HR7596, HR9087, and
HR8634 were observed for absolute flux information (Hamuy \et\
\cite{ham92}, \cite{ham94}).
\subsubsection{The \hnii\ and \sii\ line images}
\begin {figure*}
\resizebox{\hsize}{!}{\includegraphics{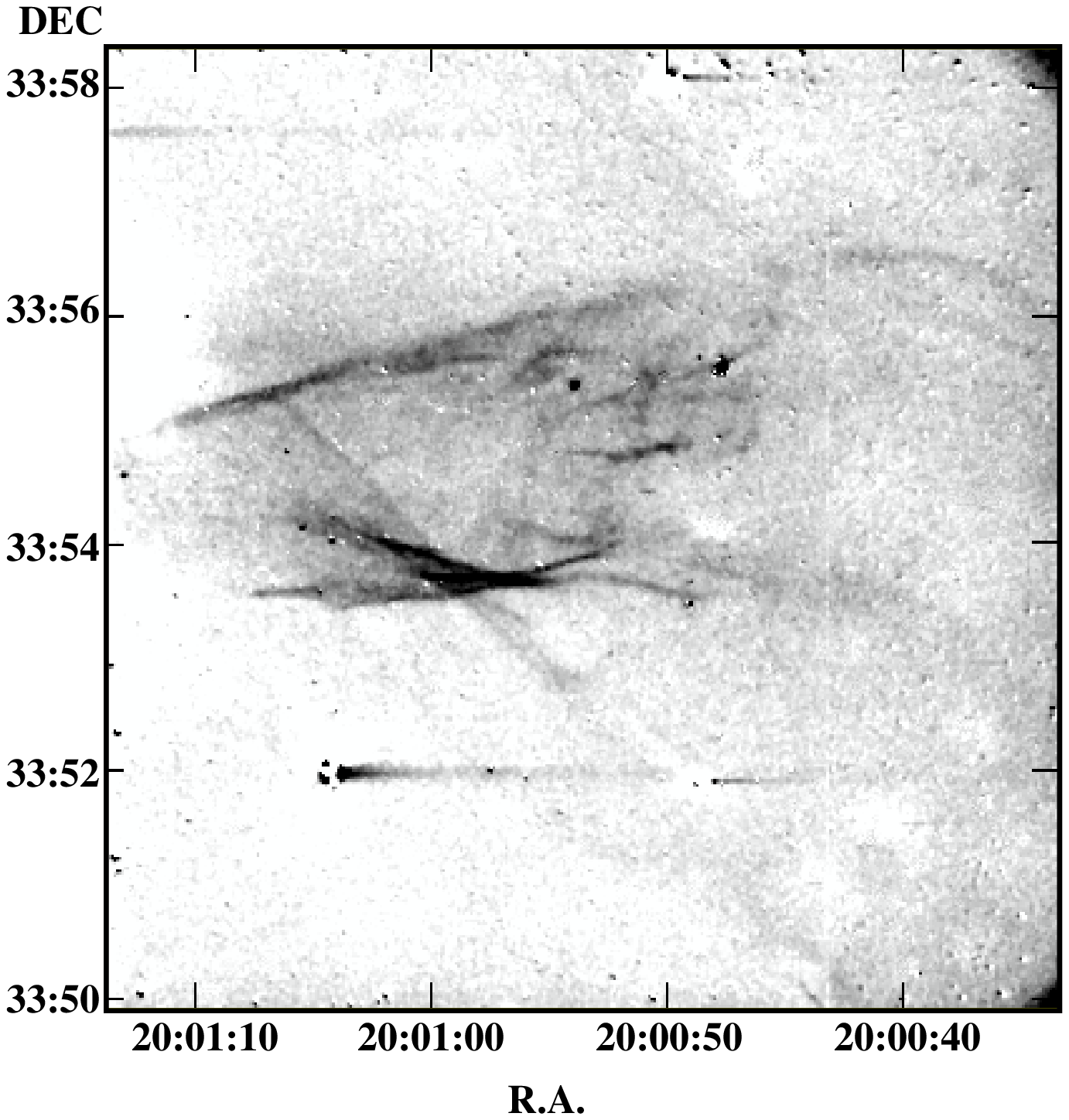}}
\caption{The 8\arcmin.5 square field covering the filamentary nebula
G70.5$+$1.9 in the light of \HNII\ after subtraction of the continuum
image.  The shadings run linearly from to 0 to 20 $\times$ \fluxa.
The line segments seen near over-exposed stars in this figure and the
following figures are due to the blooming effect.}
\label{fig01}
\end{figure*}

\begin {figure*}
\resizebox{\hsize}{!}{\includegraphics{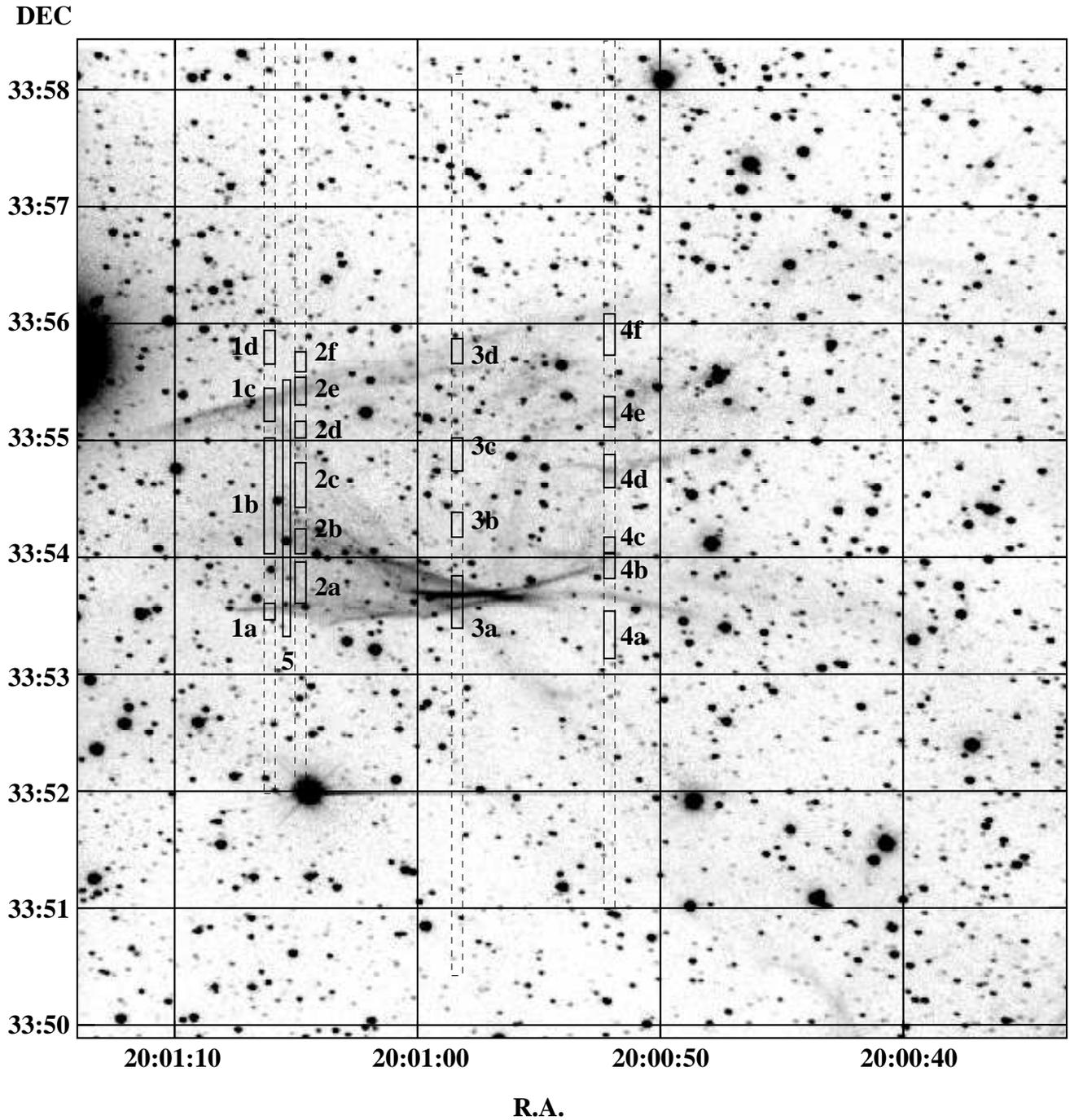}}
\caption{The image of G70.5$+$1.9 in the low
ionization line of \sii\ without continuum subtraction so that a
reference star field is clear. The long dashed rectangles show the
projection of the low--dispersion slits 1--4 on the sky, while the
small, individual rectangles along these represent the areas where
line strengths were extracted. Slit position 5 is that part of the
high--dispersion slit for which the position--velocity arrays of line
profiles are shown in Fig. 5. The shadings run linearly from to 0 to
8 $\times$ \fluxa.}
\label{fig02}
\end{figure*}

\begin {figure*}
\resizebox{\hsize}{!}{\includegraphics{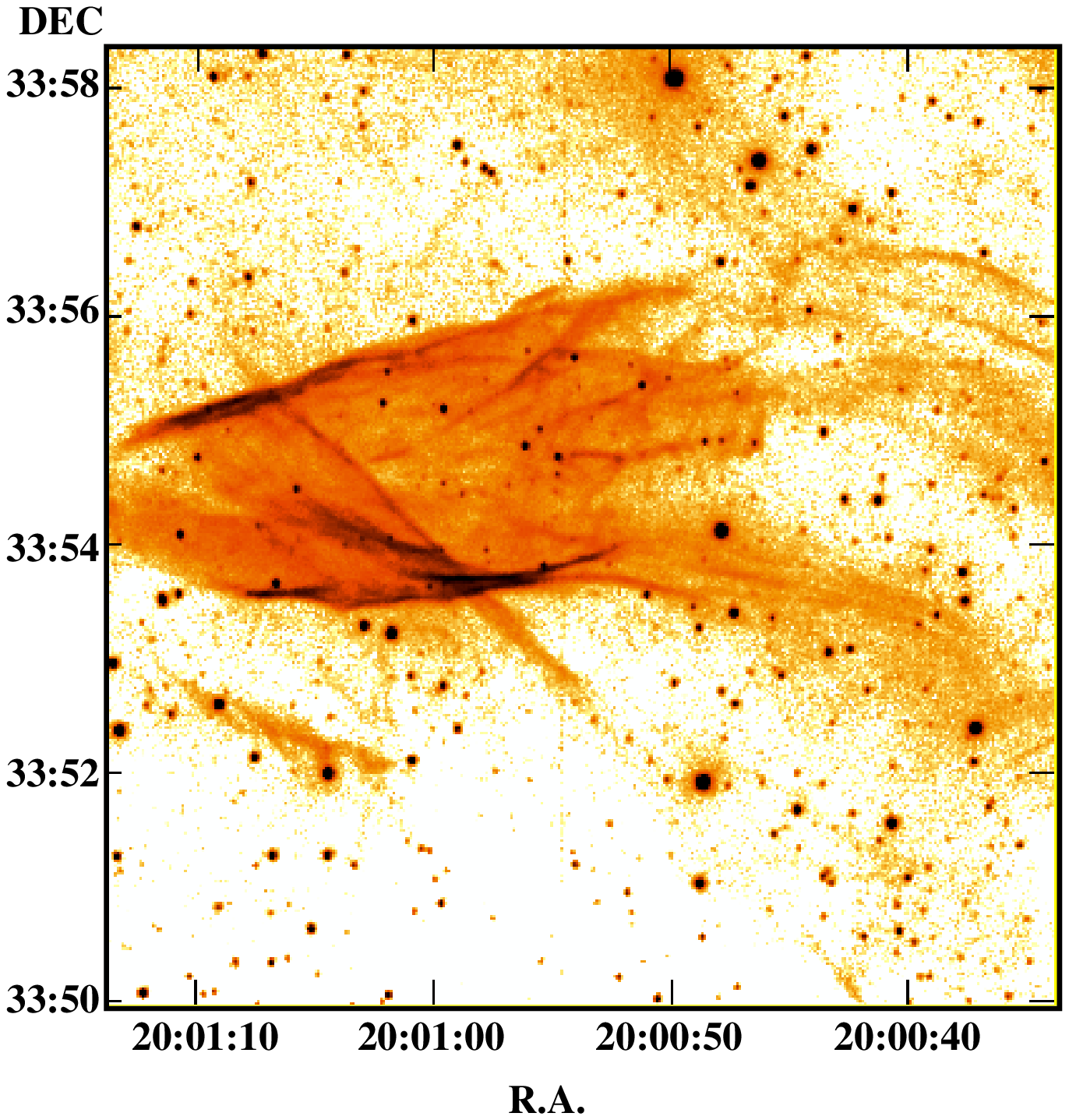}}
\caption{The 8\arcmin.5 field around G70.5$+$1.9 in the low ionization
line of \OII. The intensity representation is logarithmic and ranges
from 0 to 30 $\times$ \fluxa. The diffuse emission seen in the west
and south is at least $\sim$10 times weaker than the flux of the
bright filaments.}
\label{fig03}
\end{figure*}

The higher angular resolution images of G70.5$+$1.9 allow us to study
in more detail the network of filaments. A basic characteristic of
these filaments is their brightness in \hnii\ (Fig. \ref{fig01}) and
\sii\ (Fig. \ref{fig02}).  The bulk of the emission seems to be
bounded by two very sharp filaments.  These are separated by a typical
distance of $\sim$2\arcmin\ along the declination axis. The southern
filament is found at an almost constant declination
(\dd$\sim$33\degr53\arcmin30\arcsec), while the northern is inclined
by $\sim$15\degr\ with respect to the east--west direction. Diffuse
emission and several shorter filamentary structures are detected
between the north and south boundaries (Figs. \ref{fig01} and
\ref{fig02}). The detected emission line structures are not bounded in
the east--west direction and weak diffuse as well as filamentary
emission is detected further to west in an arc--like shape
(Fig. \ref{fig03}, \a$\simeq$20\h00\m40\s,
\dd$\simeq$33\degr53\arcmin).  There is also a long filamentary
structure originating roughly from \a$\simeq$20\h01\m07\s,
\dd$\simeq$33\degr55\arcmin20\arcsec\ and extending down to \a
$\simeq$20\h00\m42\s\ and \dd $\simeq$33\degr50\arcmin02\arcsec, after
a long gap of $\sim$3\arcmin. Finally, a patchy structure is present
around \a $\simeq$20\h01\m05\s\ and
\dd$\simeq$33\degr52\arcmin15\arcsec\ to the south of the main
emission area. The east field was also observed in these filters as
reported in Table~\ref{table1} but it is not shown here since any
emission is below our detection limit.
\subsubsection{The \oii\ and \oiii\ line mosaics}
The detected structures are also quite strong in the low ionization
line of \oii3727 \AA\ (Fig.~\ref{fig03}). The morphology in this
emission line is generally similar to that of the low ionization lines
discussed above. The south, bright filament (\a$\simeq$20\h01\m,
\dd$\simeq$33\degr53\arcmin30\arcsec) is characterized by a surface
brightness up to $\sim$ 28 $\times$ \fluxa, while the northern
filament displays a surface brightness in the range of 12--19$\times$
\fluxa.  The weak but arc--like shaped emission to the west is
detected at a level of 1--2 $\times$ \fluxa.  The typical full width
at half maximum (fwhm) of the filament in the south is found in the
range of $\sim$3\arcsec\ -- 6\arcsec, while the northern filament
(\a$\simeq$20\h01\m10\s, \dd$\simeq$33\degr55\arcmin20\arcsec) is
broader with a fwhm of $\sim$8\arcsec. The weaker filaments seen to
the interior of the bright emission are characterized by fwhm of
$\sim$3\arcsec.5.
\par
\begin {figure*}
\resizebox{\hsize}{!}{\includegraphics{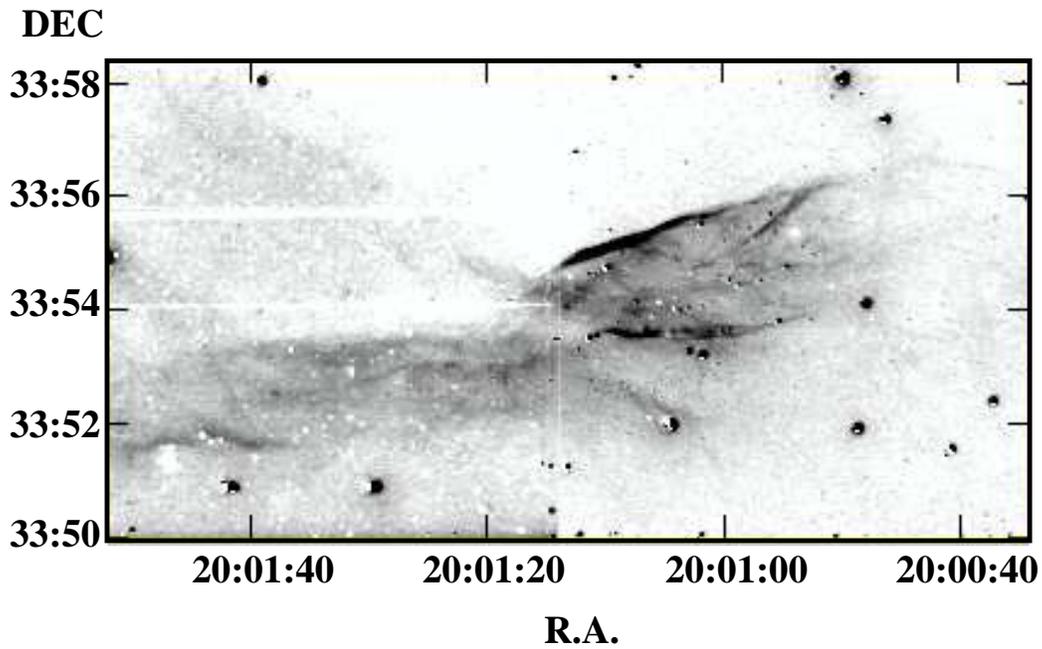}}
\caption{The \oiii\ image of the whole network of filaments after
subtraction of the continuum image. The north and south filamentary
boundaries are also prominent in the medium ionization line of \oiii
5007 \AA. However, this line reveals diffuse emission further to the
east which is not detected in the lower ionization images. The shading
run linearly from 0 to 5 $\times$ \fluxa.}
\label{fig04}
\end{figure*}

The morphology of the structure as recorded through the \oiii\
interference filter (Fig.~\ref{fig04}) differs from the morphology of
the low ionization images, including the \oii\ image.  The variety of
structures detected in the \hnii, \sii\ and \oii\ images between the
south and north filaments are not clearly present in the medium
ionization line of \oiii. In fact, some structures may not be present
at all or their morphology has now become so diffuse that does not
allow the direct identification of the corresponding structures seen
in the low ionization images.  The north boundary filament appears
quite sharp in this emission line, while it appears less well defined
in \hnii. The inverse holds in the south--west, i.e. the filament is
sharp in the \hnii\ image but diffuse emission is seen in the \oiii\
image.  A major difference between the low and medium ionization line
images is the greater extent to the east of the north and south
filaments (Fig.~\ref{fig04}).  The north \oiii\ filament extends for
$\sim$43\arcsec\ further to the east from the tip of the corresponding
\hnii\ filament, while the south \oiii\ filament extends for
$\sim$34\arcsec\ further to the east compared to the \hnii\ emission
(around 20\h01\m15\s).
The images of these regions provide direct evidence for the presence
of incomplete shock structures, i.e. areas where part of the
recombination zone is missing allowing high \oiii/\hbeta\ ratios.
Finally, the \oiii\ image displays a ``tail'' of, mainly, diffuse
emission further to the east for $\sim$8\arcmin.  The typical width of
this diffuse emission is less than $\sim$2\arcmin, while its typical
brightness is $\sim$1.5--2.0$\times$\fluxa. This diffuse component is
not detected in the low ionization images and it is not clear if it is
related to the filamentary emission in the west.
\subsection{Spectroscopy}
\subsubsection{Low dispersion}
Long--slit low--dispersion spectra were obtained on June 22 and 23,
2001 using the 1.3 m Ritchey--Cretien telescope at Skinakas
Observatory.  The data were taken with a 1300 line mm$^{-1}$ grating
and a 800 $\times$ 2000 SITe CCD covering the range of 4750 \AA\ --
6815 \AA.  The slit had a width of 7\farcs7 and a length of 7\arcmin.9
and, in all cases, was oriented in the south--north direction. The
coordinates of the slit centers along with the number of spectra and
the total exposure times are given in Table \ref{table1}.  A spectrum
taken on August 3, 2000 with the same hardware configuration and
presented by Mavromatakis \& Strom (\cite{mav02}) is also used here
with different apertures extracted.  The spectrophotometric standard
stars HR5501, HR7596, HR7950, HR9087, and HR718 were observed for
absolute flux information (Hamuy \et\ \cite{ham92}, \cite{ham94}).
\par
Long--slit spectra have been obtained at four different positions.
The relatively small size of the source (3\arcmin) compared with the
slit length of 7\arcmin.9 allowed very good background subtraction. In
addition, we have the possibility to study the nature of the diffuse
emission detected between the bright filaments. The absolute \ha\ flux
ranges from 0.5 to 7 $\times$ \fluxa\ (Table~\ref{table2}). The high
\sii/\ha\ ratios clearly demonstrate that the detected emission
originates from shock heated gas (\sii/\ha\ $\sim$0.5--1.6;
Table~\ref{table2}). The fluxes of the individual sulfur lines may
provide data on the electron density of the emitting gas (Osterbrock
\& Ferland \cite{ost06}). However, all sulfur line ratios approach the
high end of the allowable range of values suggesting low electron
densities. Using the nebular package within the IRAF software (Shaw
and Dufour \cite{sha95}) we find that all electron densities are
estimated to lie below $\sim$120 \dens. Therefore we cannot directly
estimate the preshock cloud densities but can only place upper limits,
provided that there is no magnetic field to halt the compression.

The log(\ha/\nii) versus log(\ha/\sii) intensities, from
Table~\ref{table2}, corrected for interstellar extinction, are
compared with those of other well--defined phenomena in
Fig.~\ref{fig06} (following Sabbadin et al. \cite{sab77} \&
Cant{\'{o}} \cite{can81}).
\subsubsection{High dispersion}
Observations of G70.5$+$1.9 were made with the Manchester Echelle
Spectrometer (MES--SPM -- see Meaburn et al. \cite{mea84};
\cite{mea03}) combined with the 2.1--m San Pedro Martir telescope on
23 August, 2005. A SITe CCD was the detector with 1024$\times$1024,
24~$\mu$m pixels although 2$\times$2 binning was employed throughout
the observations.
\par 
Spatially resolved, long-slit \ha, \nii\ \& \oiii\ line profiles were
obtained with the MES--SPM along the line marked 5 in
Fig.~\ref{fig02}. This is only a partial length of the full NS slits
as relevant emission only occurred over small sections of the full
slit length.  The increments along the slit length each corresponds to
0\arcsec.63.  \par In this spectroscopic mode MES--SPM has no
cross--dispersion consequently, for the present observations, a filter
of 90~\AA\ bandwidth was used to isolate the 87$^{\rm th}$ echelle
order containing the \ha\ and \nii\ nebular emission lines and one of
60~\AA\ bandwidth for \oiii. The slit width was always 150 $\mu$m
which is $\equiv$ 1.9\arcsec\ on the sky and 10 km s$^{-1}$ spectral
halfwidth.  Each integration time was 1800 s.  \par The longslit
spectra were cleaned of cosmic rays and calibrated in wavelength to
$\pm$ 1 km s$^{-1}$ accuracy in the usual way using STARLINK
\textsc{figaro} software. The greyscale representation of the
position--velocity (pv) arrays of \ha, \nii\ \& \oiii\ line profiles
for the partial slit length shown in Fig.~\ref{fig02} for Slit 5 are
shown in Fig.~\ref{fig05}. As no standard star was observed the
absolute surface brightnesses are unreliable and will not be presented
here.
\section{Discussion}
Detailed optical observations have been performed in an attempt to
understand the nature of G70.5$+$1.9. The lower ionization images in
\hnii, \sii, \oii\ and the higher ionization image in \oiii\ reveal
several filamentary structures. In addition, velocity resolved
profiles have been obtained along a specific slit position. The
current data point to a shock heated origin of the optical emission.
\par
The SNR origin of the proposed candidate remnant is strongly suggested
by the positions of the line ratios in Fig.~\ref{fig06} compared with
those of Herbig--Haro objects (HH--objects), H{\sc ii} regions and
planetary nebulae (PNe). They follow closely the shape of those
observed for those of shock ionized evolved SNRs.  The \oiii/\hbeta\
ratio is a very usefull diagnostic tool for complete or incomplete
shock structures (Raymond \et\ \cite{ray88}). All spectra suggest
complete shock structures except that from Area 1a with an
\oiii/\hbeta\ ratio of $\sim$26. Typically, this ratio is below
$\sim$6 (Cox \& Raymond \cite{cox85}; Hartigan \et\
~\cite{har87}). However, this limit is easily exceeded in case of
shocks with incomplete recombination zones like in Area 1a. Since the
long--slit spectra do not cover the full extent of the source, we can
use our flux calibrated images to map areas with incomplete shock
structures. Since our \hnii\ filter transmits equally the \nii 6548,
6584 \AA, and \ha\ lines we can estimate the \ha\ flux as $\sim$1/2 of
the flux measured in the \hnii\ filter.  Assuming that the \ha/\hbeta\
ratio is $\sim$4 all over this source, as the long--slit spectra
suggest, we estimate the \oiii/\hbeta\ ratio as $\sim$ 8 \oiii/\hnii.
Interestingly, we find that the area further to the east, i.e. between
\a\ $\sim$ 20\h01\m08\s\ and 20\h01\m14\s\ is dominated by such
structures.
\par
\begin {figure*}
\resizebox{\hsize}{!}{\includegraphics{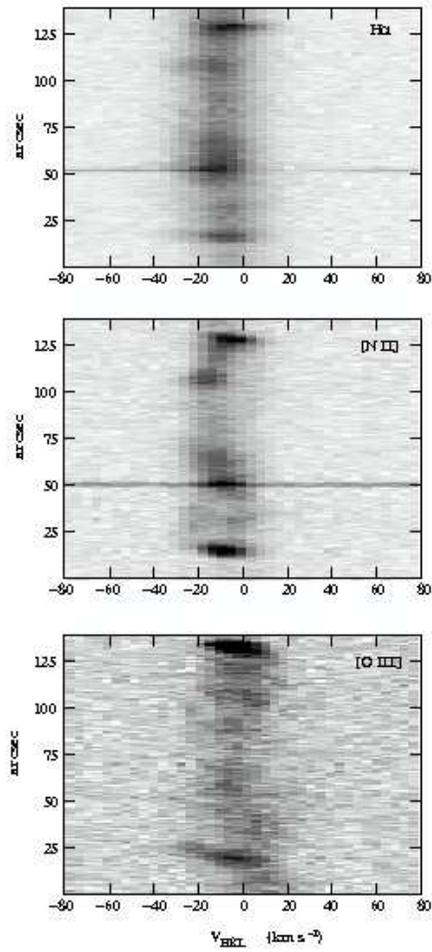}}
\caption{Greyscale representations of the position--velocity arrays of
the \ha, \nii\ \& \oiii\ line profiles from that part of the length of
slit 5 as shown in Fig.~\ref{fig02}.}
\label{fig05}
\end{figure*}

\begin {figure*}
\resizebox{\hsize}{!}{\includegraphics{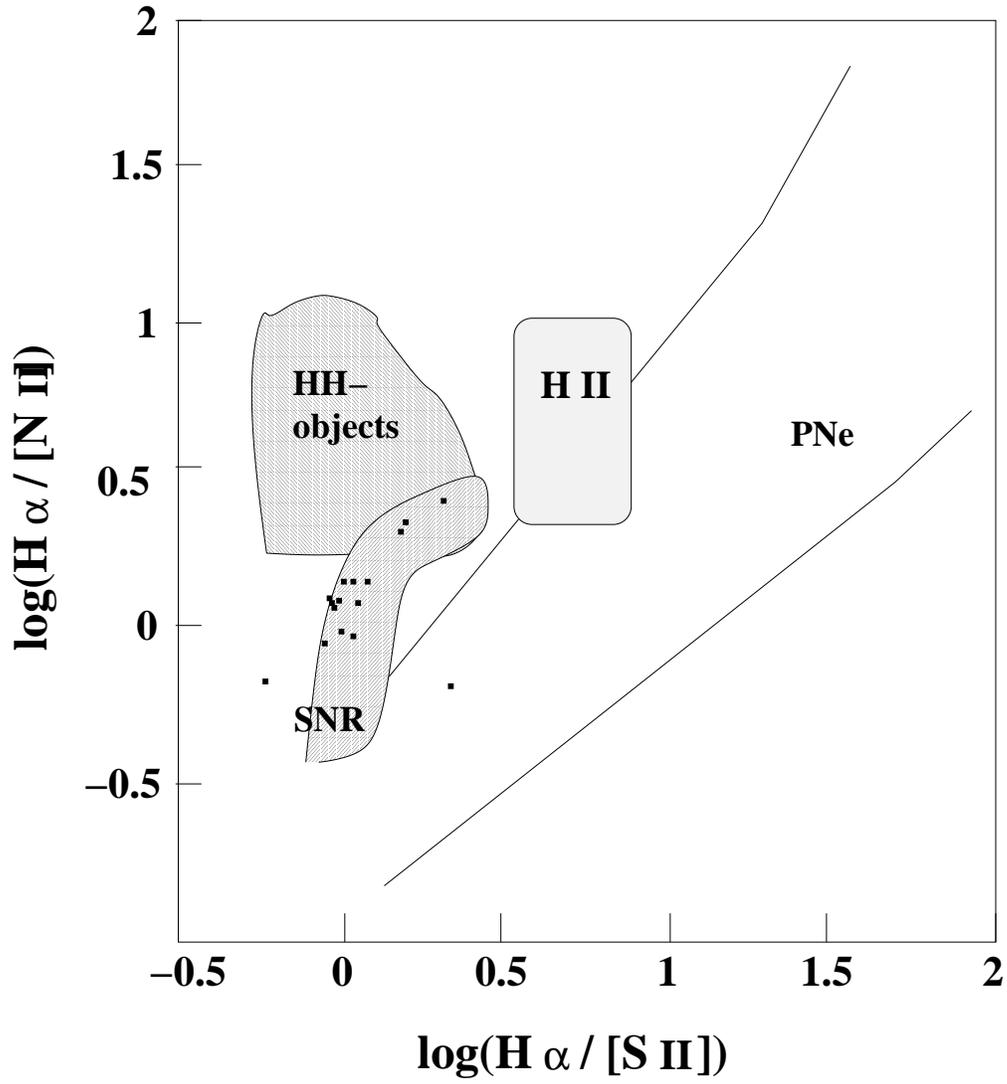}}
\caption{Diagnostic diagram (Sabbadin et al. \cite{sab77}; Cant{\'{o}}
\cite{can81}), where the positions of line ratios listed in
Table~\ref{table2}, from Areas 1a to 4f, are shown as black
squares.}
\label{fig06}
\end{figure*}

 
The motions of the filaments as seen in Fig.~\ref{fig05} also suggest
an evolved SNR origin. The line profiles are single over the bright
filaments, where expansive motions are expected to be tangential along
the line of sight, but become split, albeit by only a few tens of km
s$^{-1}$ towards the fainter regions.
\par
It remains possible that this isolated, and irregular group of
filaments is part of a wider structure but being seen through holes in
intervening clouds leading to patchy optical interstellar
extinction. The correlation of these filaments with the
structures reported by Mavromatakis \& Strom (\cite{kal76}) is also
not clear. The current data alone are not sufficient to claim a
correlation. The \ha/\hbeta\ ratios in Table 2 can be used to
estimate the variations in logarithmic extinction coefficient c over
this source assuming an intrinsic ratio of 3 and the interstellar
extinction curve of Kaler (\cite{kal76}) as implemented in the nebular
package (Shaw and Dufour (\cite{sha95}) within the IRAF software.  The
signal to noise weighted average of the observed \ha/\hbeta\ ratios in
Table \ref{table2} is 4.0 ($\pm$0.1) derived from all different
apertures and spectra available.  The logarithmic extinction c is then
0.40 which is equivalent to an A$_{\rm V}$ of 0.8 and an E(B$-$V) of
0.27. Here we have assumed E(B$-$V) = 0.664 c (Kaler \cite{kal76};
Aller \cite{all84}). E(B-V) values are listed in Table
\ref{table2} along with other parameters, where it can be seen that
statistical significant \ha/\hbeta\ ratios vary from 2.9 (area 4f) to
4.5 (area 4d).
\par
The average hydrogen column density derived from the statistical
relation of Predehl\& Schmidt (\cite{pre95}) is 1.4 $\times$ 10$^{21}$
\dens, while the total galactic hydrogen column density in the
direction of the candidate remnant is around 1$\times$ 10$^{22}$ \dens
(Kalberla et al. \cite{kal05} and Dickey \& Lockman \cite{dic90}). It
is clear that the value based on the optical data and a statistical
relation is lower by a factor of $\sim$6 than the estimated total
N{\sc h}. This implies that the detected structures are closer to us
than the total distance to the outer part of the galaxy in that
specific direction.  The use of the code given by Hakkila et
al. (\cite{hak97}) to calculate the visual extinction in the direction
of G70.5$+$1.9 supports the above suggestion given our measurements of
optical extintion (with typical color excess of 0.27). It is likely
that the distance to the proposed candidate remnant is less than 1
kpc.
\par
A crucial issue concerns the existence of radio emission. Radio
emission in the area of the optical structures is detected in the low
resolution (7\arcmin) 4850 MHz images of the Green Bank survey
(Gregory \& Condon \cite{gre91}). Given the low resolution, it is very
hard to state any spatial correlation. We have also examined the
higher resolution CGPS data at 1420 MHz (Taylor et al. \cite{tay03})
but no prominent emission was detected. The 3$\sigma$ upper limit
obtained is $\sim$4 mJy/beam.  We note here that there are also other
SNRs that do not display radio emission, at least, at the detection
level of the corresponding observations (e.g. Dickel et
al. \cite{dic01}, Filipovi$\acute{\rm c}$ et al. \cite{fil08}).  Radio
observations in different wavelengths should be performed to provide
conclusive evidence on the nature of the source as an evolved SNR.
Finally, we have searched ROSAT data for X-rays but none have been detected 
in the area of the optical emission.  
\begin{acknowledgements}
We would like to thank the referee Prof. Dickel J. for his comments
and S. Akras for his help on spectral fluxes calculations. Skinakas
Observatory is a collaborative project of the University of Crete, the
Foundation for Research and Technology-Hellas and the
Max-Planck-Institut f\"ur Extraterrestrische Physik. This research has
made use of data obtained through the High Energy Astrophysics Science
Archive Research Center Online Service, provided by the NASA/Goddard
Space Flight Center.
\end{acknowledgements}
%

%
%
\begin{table*}  
\caption[]{Imaging and Spectral log}  
\label{table1}
\begin{flushleft} 
\begin{tabular}{lcccccc}  
\noalign{\smallskip}  
\hline  
\multicolumn{6}{c}{IMAGING} \\  
\hline
Filter & $\lambda_{\rm c}$ & $\Delta \lambda$ & Total exp. time &
Fields$^{\rm a}$ \\
 & ($\AA$) & ($\AA$) & (sec) & \\
\hline
\HNII & 6566 & 75 & 6600 (3)$^{\rm b}$ & W &  \\
\SII & 6712 & 20 & 6600 (3) & W            &  \\
\OIII & 5014 & 28 & 4800 (2) & W           &  \\
\OII & 3727 & 25 & 4800 (2) & W            &  \\
Cont blue & 5470 & 230 & 360 (3) & W       &  \\
Cont red & 6096 & 134 & 510 (5) & W   	        \\
\HNII & 6566 & 75 & 2400 (1) & E 	        \\
\SII & 6712 & 20 & 2400 (1) & E 	        \\
\OIII & 5014 & 28 & 4800 (2) & E 	        \\
\OII & 3727 & 25 & 2400 (1) & E 	        \\
Cont blue & 5470 & 230 & 360 (3) & E 	        \\
Cont red & 6096 & 134 & 360 (3) & E 	        \\
\hline
\multicolumn{7}{c}{SPECTROSCOPY} \\  
\hline  
Area & \multicolumn{2}{c}{Slit centers} & Total exp. time &  Telescope& & \\  
     & $\alpha$                         & $\delta$        & (sec) &  &\\  
\hline  
Area 1 & 20\h01\m06.2\s & 33\degr56\arcmin20\arcsec &2700 (1)&  1.3-m \\
Area 2 & 20\h01\m05.0\s & 33\degr56\arcmin15\arcsec &7200 (2)&  1.3-m\\
Area 3 & 20\h00\m58.2\s & 33\degr54\arcmin34\arcsec &3600 (2)&  1.3-m\\
Area 4 & 20\h00\m52.0\s & 33\degr55\arcmin41\arcsec &3600 (2)&  1.3-m\\
Area 5 & 20\h01\m05.4\s & 33\degr54\arcmin20\arcsec &3600 (2)&  2.1-m\\
\hline  
\end{tabular}
\end{flushleft}
\begin{flushleft}
${\rm ^a}$ Field observed: W($=$West), E($=$East).\\ 
${\rm ^b}$ Numbers in parentheses represent the number of individual
frames.\\
\end{flushleft} 
\end{table*}  
%
\begin{table*}
\caption[]{Relative line fluxes.}
\label{table2}
\begin{flushleft}
\begin{tabular}{llllllllllllllll}
\hline \noalign{\smallskip} & \multicolumn{3}{c}{Area 1a} &
\multicolumn{3}{c}{Area 1b} & \multicolumn{3}{ c}{Area 1c} &
\multicolumn{3}{ c}{Area 1d} & \multicolumn{3}{ c}{Area 2a} \\
Line (\AA) & F$^{\rm a}$ & I$^{\rm b}$ & S/N$^{\rm c}$ & F & I & S/N
&F & I & S/N & F & I & S/N & F & I & S/N \\
\hline \hbeta\ 4861 & 15 & 36 & (6) & 26 & 33 & (43) & 24 & 34 & (26)&
20 & 34 & (11) & 24 & 34 & (28) \\
\oxygen\ 4959 & 99 & 226 & (39) & 35 & 44 & (61) & 32 & 45 & (40) &
--& -- & -- & 31 & 43 & (39) \\
\oxygen\ 5007 & 317 & 702 & (114) & 107 & 134 & (177) & 103 & 142
&(115) & -- & -- & -- & 92 & 126 & (111) \\
\oi\ 6300 & 11 & 11 & (7) & 5 & 5 & (17) & -- & -- & -- & -- & -- &
--& 7 & 7 & (10) \\
\oi\ 6364 & -- & -- & -- & -- & -- & -- & -- & -- & -- & -- & -- & --&
2 & 2 & (3) \\
\nitrogen\ 6548 & 40 & 40 & (27) & 25 & 25 & (68) & 25 & 25 & (50) &15
& 15 & (21) & 27 & 27 & (54) \\
\ha\ 6563 & 100 & 100 & (47) & 100 & 100 & (158) & 100 & 100 & (133)
&100 & 100 & (105) & 100 & 100 & (147) \\
\nitrogen\ 6583 & 141 & 140 & (72) & 79 & 79 & (148) & 84 & 84 & (35)&
47 & 47 & (53) & 89 & 89 & (144) \\
\sulfur\ 6716 & 91 & 86 & (55) & 50 & 49 & (119) & 51 & 50 & (87) &
29& 28 & (41) & 68 & 66 & (125) \\
\sulfur\ 6731 & 68 & 64 & (40) & 36 & 35 & (76) & 37 & 36 & (67) & 21&
20 & (23) & 49 & 48 & (94) \\
\hline Absolute \ha\ flux$^{\rm d}$ & \multicolumn{3}{c}{2} &
\multicolumn{3}{c}{2.1} & \multicolumn{3}{c}{3.9} &
\multicolumn{3}{c}{1.3} & \multicolumn{3}{c}{2.5} \\
\sulfur/\ha\ & \multicolumn{3}{c}{1.50 $\pm$ 0.04}
&\multicolumn{3}{c}{0.84 $\pm$ 0.01} & \multicolumn{3}{c}{0.86 $\pm$
  0.01} & \multicolumn{3}{c}{0.48 $\pm$ 0.01} &
\multicolumn{3}{c}{1.14$\pm$ 0.01} \\
F(6716)/F(6731) & \multicolumn{3}{c}{1.34 $\pm$ 0.04}
&\multicolumn{3}{c}{1.39 $\pm$ 0.03} & \multicolumn{3}{c}{1.38 $\pm$
  0.03} & \multicolumn{3}{c}{1.38 $\pm$ 0.07} &
\multicolumn{3}{c}{1.39$\pm$ 0.02} \\
\nii/\ha\ & \multicolumn{3}{c}{1.80 $\pm$ 0.05}
&\multicolumn{3}{c}{1.04 $\pm$ 0.02} & \multicolumn{3}{c}{1.09 $\pm$
  0.01} & \multicolumn{3}{c}{0.62 $\pm$ 0.01} &
\multicolumn{3}{c}{1.16$\pm$ 0.01} \\
\oxygen/\hbeta\ & \multicolumn{3}{c}{25.8 $\pm$ 4.3} &
\multicolumn{3}{c}{5.40 $\pm$ 0.13} & \multicolumn{3}{c}{5.50 $\pm$
0.21} & \multicolumn{3}{c}{--} & \multicolumn{3}{c}{5.00 $\pm$ 0.18}\\
c(\hbeta) & \multicolumn{3}{c}{1.10 $\pm$ 0.20} &
\multicolumn{3}{c}{0.31 $\pm$ 0.03} & \multicolumn{3}{c}{0.44 $\pm$
0.05} & \multicolumn{3}{c}{0.67 $\pm$ 0.12} & \multicolumn{3}{c}{0.44
$\pm$ 0.05} \\
E$_{\rm B-V}$ & \multicolumn{3}{c}{0.73 $\pm$ 0.13}
&\multicolumn{3}{c}{0.21 $\pm$ 0.02} & \multicolumn{3}{c}{0.29 $\pm$
  0.03} & \multicolumn{3}{c}{0.44 $\pm$ 0.08} &
\multicolumn{3}{c}{0.29$\pm$ 0.03} \\
\hline 
\noalign{\smallskip} & \multicolumn{3}{c}{Area 2b} &
\multicolumn{3}{c}{Area 2c} & \multicolumn{3}{ c}{Area 2d} &
\multicolumn{3}{ c}{Area 2e} & \multicolumn{3}{ c}{Area 2f} \\
Line (\AA) & F$^{\rm a}$ & I$^{\rm b}$ & S/N$^{\rm c}$ & F & I & S/N
&F & I & S/N & F & I & S/N & F & I & S/N \\
\hline \hbeta\ 4861 & 24 & 34 & (35) & 23 & 34 & (25) & 25 & 34 &
(21)& 24 & 34 & (38) & 29 & 33 & (17) \\
\oxygen\ 4959 & 18 & 25 & (30) & 25 & 36 & (30) & 32 & 43 & (29) & 39&
54 & (69) & 5 & 6 & (2) \\
\oxygen\ 5007 & 56 & 77 & (88) & 72 & 101 & (83) & 97 & 128 &(87) &
119 & 164 & (190) & 12 & 14 & (9) \\
\oi\ 6300 & 9 & 9 & (17) & 7 & 7 & (9) & 10 & 10 & (10) & 6 & 6 &
(12)& 10 & 10 & (5) \\
\oi\ 6364 & 2 & 2 & (5) & 3 & 3 & (5) & 3 & 3 & (5) & -- & -- & -- &--
& -- & -- \\
\nitrogen\ 6548 & 24 & 24 & (60) & 21 & 21 & (41) & 23 & 23 & (36) &26
& 26 & (72) & 17 & 17 & (19) \\
\ha\ 6563 & 100 & 100 & (174) & 100 & 100 & (140) & 100 & 100 & (109)
&100 & 100 & (197) & 100 & 100 & (77) \\
\nitrogen\ 6583 & 78 & 78 & (159) & 69 & 68 & (111) & 71 & 71 & (89)&
84 & 84 & (189) & 48 & 48 & (44) \\
\sulfur\ 6716 & 63 & 62 & (159) & 51 & 50 & (92) & 44 & 43 & (64) &
50& 49 & (130) & 30 & 30 & (44) \\
\sulfur\ 6731 & 45 & 44 & (102) & 36 & 35 & (69) & 32 & 31 & (49) &
36& 35 & (99) & 22 & 22 & (25) \\
\hline Absolute \ha\ flux$^{\rm d}$ & \multicolumn{3}{c}{4.2} &
\multicolumn{3}{c}{2.1} & \multicolumn{3}{c}{2.2} &
\multicolumn{3}{c}{3.7} & \multicolumn{3}{c}{1.5} \\
\sulfur/\ha\ & \multicolumn{3}{c}{1.06 $\pm$ 0.01} &
\multicolumn{3}{c}{0.85 $\pm$ 0.01} & \multicolumn{3}{c}{0.74 $\pm$
0.01} & \multicolumn{3}{c}{0.84 $\pm$ 0.01} & \multicolumn{3}{c}{0.52
$\pm$ 0.01} \\
F(6716)/F(6731) & \multicolumn{3}{c}{1.40 $\pm$ 0.02} &
\multicolumn{3}{c}{1.42 $\pm$ 0.03} & \multicolumn{3}{c}{1.38 $\pm$
0.04} & \multicolumn{3}{c}{1.39 $\pm$ 0.02} & \multicolumn{3}{c}{1.36
$\pm$ 0.07} \\
\nii/\ha\ & \multicolumn{3}{c}{1.02 $\pm$ 0.01} &
\multicolumn{3}{c}{0.89 $\pm$ 0.01} & \multicolumn{3}{c}{0.94 $\pm$
0.01} & \multicolumn{3}{c}{1.10 $\pm$ 0.01} & \multicolumn{3}{c}{0.65
$\pm$ 0.01} \\
\oxygen/\hbeta\ & \multicolumn{3}{c}{3.00 $\pm$ 0.09} &
\multicolumn{3}{c}{4.00 $\pm$ 0.18} & \multicolumn{3}{c}{5.00 $\pm$
0.25} & \multicolumn{3}{c}{6.40 $\pm$ 0.17} & \multicolumn{3}{c}{0.60
$\pm$ 0.07}\\
c(\hbeta) & \multicolumn{3}{c}{0.44 $\pm$ 0.04} &
\multicolumn{3}{c}{0.47 $\pm$ 0.05} & \multicolumn{3}{c}{0.38 $\pm$
0.07} & \multicolumn{3}{c}{0.44 $\pm$ 0.04} & \multicolumn{3}{c}{0.16
$\pm$ 0.08} \\
E$_{\rm B-V}$ & \multicolumn{3}{c}{0.29 $\pm$ 0.03} &
\multicolumn{3}{c}{0.31 $\pm$ 0.03} & \multicolumn{3}{c}{0.25 $\pm$
0.05} & \multicolumn{3}{c}{0.29 $\pm$ 0.03} & \multicolumn{3}{c}{0.11
$\pm$ 0.06} \\
\hline
\end{tabular}
\end{flushleft}  
\begin{flushleft}

${\rm ^a}$ Observed surface brightness normalized to F(H$\alpha$)=100 and
uncorrected for interstellar extinction.\\

${\rm ^b}$ Intrinsic surface brightness normalized to F(H$\alpha$)=100 and
corrected for interstellar extinction.\\

${\rm ^c}$ Numbers in parentheses represent the signal--to--noise
ratio of the quoted fluxes.\\

$^{\rm d}$ In units of \fluxa.\\  

The letters next to the areas number indicate different apertures
extracted along the slit.\\

Listed fluxes are a signal--to--noise weighted average of two fluxes
for areas 2, 3 and 4.\\

The emission line ratios  
\sulfur/\ha, F(6716)/F(6731), \nii/\ha\ and \oxygen/\hbeta\ are
calculated using the values corrected for interstellar extinction.\\

The errors of the emission line ratios, c(\hbeta) and E$_{\rm B-V}$,
are calculated through standard error propagation.\\

\end{flushleft}   
\end{table*}
%
\begin{table*}
\caption[]{Table 2 (continued)}
\label{table3}
\begin{flushleft}
\begin{tabular}{llllllllllllllll}
\hline \noalign{\smallskip} & \multicolumn{3}{c}{Area 3a} &
\multicolumn{3}{c}{Area 3b} & \multicolumn{3}{ c}{Area 3c} &
\multicolumn{3}{ c}{Area 3d} & \multicolumn{3}{ c}{Area 4a} \\
Line (\AA) & F$^{\rm a}$ & I$^{\rm b}$ & S/N$^{\rm c}$ & F & I & S/N &
F & I & S/N & F & I & S/N & F & I & S/N \\
\hline \hbeta\ 4861 & 25 & 34 & (83) & 28 & 34 & (12) & 26 & 33 & (19)
& 25 & 34 & (23) & 27 & 34 & (9) \\
\oxygen\ 4959 & 9 & 12 & (33) & 26 & 31 & (12) & 32 & 40 & (24) & 14
& 19 & (15) & 11 & 14 & (3) \\
\oxygen\ 5007 & 25 & 33 & (89) & 84 & 100 & (37) & 109 & 136 &
(80) & 46 & 61 & (48) & 57 & 70 & (20) \\
\oi\ 6300 & 7 & 7 & (40) & -- & -- & -- & -- & -- & -- & -- & -- & --
& 50 & 51 & (10) \\
\oi\ 6364 & 3 & 3 & (18) & -- & -- & -- & -- & -- & -- & -- & -- & --
& 17 & 17 & (5) \\
\nitrogen\ 6548 & 22 & 22 & (127) & 27 & 27 & (20) & 18 & 18 & (23) &
24 & 24 & (40) & -- & -- & -- \\
\ha\ 6563 & 100 & 100 & (383) & 100 & 100 & (59) & 100 & 100 & (98) &
100 & 100 & (123) & 100 & 100 & (44) \\
\nitrogen\ 6583 & 72 & 72 & (317) & 82 & 82 & (53) & 65 & 65 & (72)
& 76 & 76 & (107) & 43 & 43 & (17) \\
\sulfur\ 6716 & 65 & 64 & (301) & 55 & 54 & (39) & 45 & 44 & (55) & 44
& 43 & (67) & 95 & 94 & (42) \\
\sulfur\ 6731 & 46 & 45 & (218) & 36 & 35 & (27) & 31 & 30 & (38) & 32
& 31 & (52) & 64 & 63 & (31) \\
\hline Absolute \ha\ flux$^{\rm d}$ & \multicolumn{3}{c}{6.9} &
\multicolumn{3}{c}{1.6} & \multicolumn{3}{c}{1.8} &
\multicolumn{3}{c}{3.7} & \multicolumn{3}{c}{0.5} \\
\sulfur/\ha\ & \multicolumn{3}{c}{1.09 $\pm$ 0.01} &
\multicolumn{3}{c}{0.89 $\pm$ 0.02} & \multicolumn{3}{c}{0.74 $\pm$
0.01} & \multicolumn{3}{c}{0.74 $\pm$ 0.01} & \multicolumn{3}{c}{1.57
$\pm$ 0.05} \\
F(6716)/F(6731) & \multicolumn{3}{c}{1.41 $\pm$ 0.01} &
\multicolumn{3}{c}{1.53 $\pm$ 0.07} & \multicolumn{3}{c}{1.45 $\pm$
0.05} & \multicolumn{3}{c}{1.37 $\pm$ 0.03} & \multicolumn{3}{c}{1.48
$\pm$ 0.06} \\
\nii/\ha\ & \multicolumn{3}{c}{0.94 $\pm$ 0.01} &
\multicolumn{3}{c}{1.09 $\pm$ 0.02} & \multicolumn{3}{c}{0.83 $\pm$
0.01} & \multicolumn{3}{c}{1.00 $\pm$ 0.01} & \multicolumn{3}{c}{0.43
$\pm$ 0.01} \\
\oxygen/\hbeta\ & \multicolumn{3}{c}{1.32 $\pm$ 0.02} &
\multicolumn{3}{c}{3.80 $\pm$ 0.32} & \multicolumn{3}{c}{5.30 $\pm$ 0.28}
& \multicolumn{3}{c}{2.40 $\pm$ 0.11} & \multicolumn{3}{c}{2.50 $\pm$
0.31}\\
c(\hbeta) & \multicolumn{3}{c}{0.38 $\pm$ 0.02} &
\multicolumn{3}{c}{0.24 $\pm$ 0.11} & \multicolumn{3}{c}{0.31 $\pm$
0.07} & \multicolumn{3}{c}{0.38 $\pm$ 0.06} & \multicolumn{3}{c}{0.28
$\pm$ 0.15} \\
E$_{\rm B-V}$ & \multicolumn{3}{c}{0.25 $\pm$ 0.01} &
\multicolumn{3}{c}{0.16 $\pm$ 0.07} & \multicolumn{3}{c}{0.21 $\pm$
0.05} & \multicolumn{3}{c}{0.25 $\pm$ 0.04} & \multicolumn{3}{c}{0.19
$\pm$ 0.10} \\
\hline 
\noalign{\smallskip} & \multicolumn{3}{c}{Area 4b} &
\multicolumn{3}{c}{Area 4c} & \multicolumn{3}{ c}{Area 4d} &
\multicolumn{3}{ c}{Area 4e} & \multicolumn{3}{ c}{Area 4f} \\
Line (\AA) & F$^{\rm a}$ & I$^{\rm b}$ & S/N$^{\rm c}$ & F & I & S/N &
F & I & S/N & F & I & S/N & F & I & S/N \\
\hline \hbeta\ 4861 & 26 & 33 & (30) & 23 & 34 & (22) & 22 & 34 & (37)
& 28 & 34 & (30) & 35 & 36 & (29) \\
\oxygen\ 4959 & 7 & 9 & (10) & 6 & 9 & (7) & 11 & 16 & (20) & 11
& 13 & (15) & 14 & 15 & (14) \\
\oxygen\ 5007 & 35 & 44 & (46) & 37 & 52 & (39) & 50 & 73 &
(92) & 41 & 49 & (50) & 52 & 54 & (48) \\
\oi\ 6300 & -- & -- & -- & 6 & 6 & (7) & 11 & 12 & (24) & 10 & 10 & (12)
& 13 & 13 & (9) \\
\oi\ 6364 & -- & -- & -- & -- & -- & -- & 3 & 3 & (8) & 4 & 4 & (6) &
3 & 3 & (2) \\
\nitrogen\ 6548 & 28 & 28 & (59) & 20 & 20 & (34) & 29 & 29 & (86) &
24 & 24 & (46) & 18 & 18 & (27) \\
\ha\ 6563 & 100 & 100 & (156) & 100 & 100 & (118) & 100 & 100 & (215) &
100 & 100 & (144) & 100 & 100 & (109) \\
\nitrogen\ 6583 & 88 & 88 & (148) & 66 & 66 & (90) & 101 & 101 & (233)
& 80 & 80 & (127) & 57 & 57 & (73) \\
\sulfur\ 6716 & 66 & 65 & (127) & 51 & 49 & (78) & 68 & 66 & (182) & 55
& 54 & (101) & 37 & 37 & (58) \\
\sulfur\ 6731 & 48 & 47 & (100) & 34 & 33 & (56) & 49 & 47 & (131) & 41
& 40 & (81) & 26 & 26 & (42) \\
\hline Absolute \ha\ flux$^{\rm d}$ & \multicolumn{3}{c}{2.6} &
\multicolumn{3}{c}{3.4} & \multicolumn{3}{c}{2.7} &
\multicolumn{3}{c}{2.6} & \multicolumn{3}{c}{1.9} \\
\sulfur/\ha\ & \multicolumn{3}{c}{1.12 $\pm$ 0.01} &
\multicolumn{3}{c}{0.82 $\pm$ 0.01} & \multicolumn{3}{c}{1.13 $\pm$
0.01} & \multicolumn{3}{c}{0.94 $\pm$ 0.01} & \multicolumn{3}{c}{0.63
$\pm$ 0.01} \\
F(6716)/F(6731) & \multicolumn{3}{c}{1.38 $\pm$ 0.02} &
\multicolumn{3}{c}{1.48 $\pm$ 0.03} & \multicolumn{3}{c}{1.40 $\pm$
0.01} & \multicolumn{3}{c}{1.35 $\pm$ 0.02} & \multicolumn{3}{c}{1.42
$\pm$ 0.04} \\
\nii/\ha\ & \multicolumn{3}{c}{1.16 $\pm$ 0.01} &
\multicolumn{3}{c}{0.86 $\pm$ 0.01} & \multicolumn{3}{c}{1.30 $\pm$
0.01} & \multicolumn{3}{c}{1.04 $\pm$ 0.01} & \multicolumn{3}{c}{0.75
$\pm$ 0.01} \\
\oxygen/\hbeta\ & \multicolumn{3}{c}{1.60 $\pm$ 0.06} &
\multicolumn{3}{c}{1.80 $\pm$ 0.09} & \multicolumn{3}{c}{2.60 $\pm$ 0.08}
& \multicolumn{3}{c}{1.82 $\pm$ 0.07} & \multicolumn{3}{c}{1.92 $\pm$
0.08}\\
c(\hbeta) & \multicolumn{3}{c}{0.31 $\pm$ 0.04} &
\multicolumn{3}{c}{0.47 $\pm$ 0.06} & \multicolumn{3}{c}{0.53 $\pm$
0.04} & \multicolumn{3}{c}{0.24 $\pm$ 0.04} & \multicolumn{3}{c}{0.16
$\pm$ 0.05} \\
E$_{\rm B-V}$ & \multicolumn{3}{c}{0.21 $\pm$ 0.03} &
\multicolumn{3}{c}{0.31 $\pm$ 0.04} & \multicolumn{3}{c}{0.35 $\pm$
0.03} & \multicolumn{3}{c}{0.16 $\pm$ 0.03} & \multicolumn{3}{c}{0.11
$\pm$ 0.03} \\
\hline 
\end{tabular}
\end{flushleft}  
\begin{flushleft}

${\rm ^a}$ Observed surface brightness normalized to F(H$\alpha$)=100
and uncorrected for interstellar extinction.\\

${\rm ^b}$ Intrinsic surface brightness normalized to F(H$\alpha$)=100
and corrected for interstellar extinction.\\

${\rm ^c}$ Numbers in parentheses represent the signal--to--noise
ratio of the quoted fluxes.\\

$^{\rm d}$ In units of \fluxa.\\

The letters next to the areas number indicate different apertures
extracted along the slit.\\

Listed fluxes are a signal--to--noise weighted average of two fluxes
for areas 2, 3 and 4.\\

The emission line ratios 
\nii/\ha\ and \oxygen/\hbeta\ are calculated using the values
corrected for interstellar extinction.\\

\end{flushleft}   
\end{table*}
%
%
\newpage
\vfill\eject
%
%
%
%
%

\end{document}